\documentclass{PoS}

\title{Study of constant mode in charmonium correlators at finite temperature}

\ShortTitle{Study of constant mode in charmonium correlators at finite temperature}

\author{\speaker{Takashi Umeda}\\
       Graduate School of Pure and Applied Sciences, University of
       Tsukuba, Tsukuba, Ibaraki 305-8571, Japan \\ 
        E-mail: \email{tumeda@het.ph.tsukuba.ac.jp}}

\abstract{Recent studies on the spectral function of charmonium in
lattice QCD suggest survival of $J/\psi$ state in the deconfinement phase
till relatively high temperature. Based on the studies, different
scenarios of $J/\psi$ suppression are discussed to understand experimental
results in the Heavy Ion Collision experiments. The scenarios require
the information on the dissociation temperatures of $\chi_c$ and $\psi'$ as
well as that of $J/\psi$. In order to investigate these states in finite
temperature lattice QCD, we have to consider an effect of a
characteristic constant mode in the correlators. As a result of the
study on the constant mode, we find that most drastic change in
charmonium correlators for $\chi_c$ states just above the deconfinement
transition are caused by the constant mode. It may indicate the
survival of $\chi_c$ states after the deconfinement transition until, at
least, $1.4T_c$. }

\FullConference{The XXV International Symposium on Lattice Field Theory\\
		 July 30-4 August 2007\\
		 Regensburg, Germany}

\begin{document}

\section{Introduction}

Study of heavy quarkona is important to
understand the quark gluon plasma (QGP) formation in heavy ion
collision experiments e.g. the RHIC experiment at Brookhaven National
Laboratory. 
Recent studies on the spectral function of charmonium above $T_c$
suggest that hadronic excitations corresponding to $J/\psi$ may survive
in the deconfinement phase till relatively high temperature
\cite{Umeda:2000ym,Umeda:2002vr,Datta:2003ww,Asakawa:2003re,Aarts:2006nr,Jakovac:2006sf}.
Such results of strongly interacting QGP may affect the  
scenario of $J/\psi$ suppression \cite{Hashimoto:1986nn,Matsui:1986dk}.
Therefore the determination of the accurate dissociation temperature
of $J/\psi$ is required for many phenomenological studies.

\begin{figure}
\begin{center}
 \includegraphics[width=.35\textwidth]{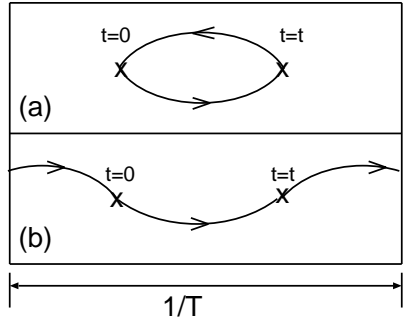}
\caption{A sketch of quark line diagrams for a meson-like correlator 
in a system with finite temporal extent. 
Vertical lines show the boundaries in temporal direction.}
\label{fig:corr}
\end{center}
\end{figure}

In order to study temporal correlators at finite temperature, 
one has to take account of a characteristic contribution caused
by the finite temporal extent.
Usually a meson correlator is interpreted by a diagram with quark and
anti-quark propagators like it is sketched in Fig.~\ref{fig:corr}(a).
However in the case of a system with a finite temporal extent, a
wraparound (scattering) contribution, as shown in
Fig.~\ref{fig:corr}(b), has to be included as well. 
The effects are also known even at zero temperature case
e.g. the study of two pion correlators \cite{Kim:2003xt}, and pentaquark 
correlator \cite{Takahashi:2005uk}.
In the case, the contribution is usually removed using e.g. 
Dirichlet boundary conditions \cite{Takahashi:2005uk} 
or appropriate analyses \cite{Kim:2003xt}.
On the other hand, at finite temperature, the latter contribution 
is genuine, and it provides a constant contribution to the correlator
as a zero energy mode. 
Due to the contribution, correlators of such meson-like operators,
e.g. charmonia, may be drastically changed just after the deconfinement
transition.

In this proceeding, we show how the constant mode affects the 
thermal effect in the charmonium correlators or the spectral functions.
Details of the study has already been reported in the 
published paper \cite{Umeda:2007hy}. 

\section{Constant mode in free quark case}
\label{sec:free}

\subsection{Meson-like correlators in the free quark case}

In order to investigate the constant contribution, 
we first consider the free quark case of QCD, in which the constant
contribution can be easily calculated.
Here we define the meson-like correlators with quark
bilinear operators,  
$O_{\Gamma}(\vec{x},t)=\bar{q}(\vec{x},t)\Gamma q(\vec{x},t)$,
in the free quark case this gives for the correlator,
\begin{eqnarray}
C(t)=\sum_{\vec{x}}\langle O_\Gamma(\vec{x},t)
O_\Gamma^\dagger(\vec{0},0) \rangle,
\label{eq:corrdef}
\end{eqnarray}
where $\Gamma$ are appropriate $4\times 4$ matrices, i.e.
$\gamma_5$, $\gamma_i$, $1$, and $\gamma_i\gamma_5$ for 
pseudoscalar(Ps), vector(V), scalar(S), and axialvector(Av) channels
respectively.
In this paper we calculate correlators for degenerate quark masses,
constructed from bilinear operators, with vanishing spatial momentum as
the simplest case, 
which is aiming at studies of charmonium at finite temperature
discussed latter.
The spectral function of the correlator is defined by
\begin{eqnarray}
C(t)&=&\int_0^\infty d\omega \rho_\Gamma(\omega)K(\omega,t), \nonumber\\
&&K(\omega,t)=\frac{\cosh\left(\omega(\frac{L_t}{2}-t)\right)}
{\sinh{\left(\omega\frac{L_t}{2}\right)}},
\label{eq:spf}
\end{eqnarray}
and has been calculated in Ref.~\cite{Karsch:2003wy,Aarts:2005hg} 
in the high temperature limit.

\subsection{Free quark calculations on a lattice}
\label{sec:freecorr} 

Here we present numerical calculations of the meson-like 
correlators for free quarks on a lattice.
The calculations are performed on an
isotropic $N_s^3\times N_t = 16^3 \times 32$ lattice.
The free quark is described by the Wilson quark action with a bare quark
mass of $m_q=0.2$. 

Figure \ref{fig:free} (left panel) shows the effective masses of the
correlators defined above.
The effective mass is defined by correlator at successive time
slices.
\begin{equation}
\frac{C(t)}{C(t+1)}=
\frac{\cosh{\left[m_{\mbox{eff}}(t)\left(\frac{N_t}{2}-t\right)\right]}}
{\cosh{\left[m_{\mbox{eff}}(t)\left(\frac{N_t}{2}-t-1\right)\right]}}
\end{equation}
The effective masses of the meson-like correlators with free quarks
should approach the energy of the two quark state without momentum for 
S-wave states (Ps and V channels) and with a minimum momentum for 
P-wave states (S and Av channels) except for the zero energy mode.
In the Ps and V channels, the zero relative momentum processes have no
constant contribution and their effective masses approach the
expected values. 
In the S and Av channels, on the other hand, their effective masses
approach the zero energy level rather than the energy of two quark
states. 

\subsection{An analysis to avoid the constant contribution}

Next we consider how to estimate the effect of the constant mode.
Of course the cosh + constant fit enables us to do so, but here we
present the method without a fit analysis.
The method uses the midpoint subtracted correlator, 
\begin{eqnarray}
 \bar{C}(t)=C(t)-C(N_t/2).
\end{eqnarray}
One can also define the effective mass of the subtracted correlator,
\begin{eqnarray}
\frac{\bar{C}(t)}{\bar{C}(t+1)}=
\frac{\sinh^2{\left[\frac{1}{2}m_{\mbox{eff}}^{\mbox{sub}}(t)
\left(\frac{N_t}{2}-t\right)\right]}}
{\sinh^2{\left[\frac{1}{2}m_{\mbox{eff}}^{\mbox{sub}}(t)
\left(\frac{N_t}{2}-t-1\right)\right]}}.
\label{eq:effmasssub}
\end{eqnarray}

\begin{figure}
\begin{center}
 \includegraphics[width=.4\textwidth]{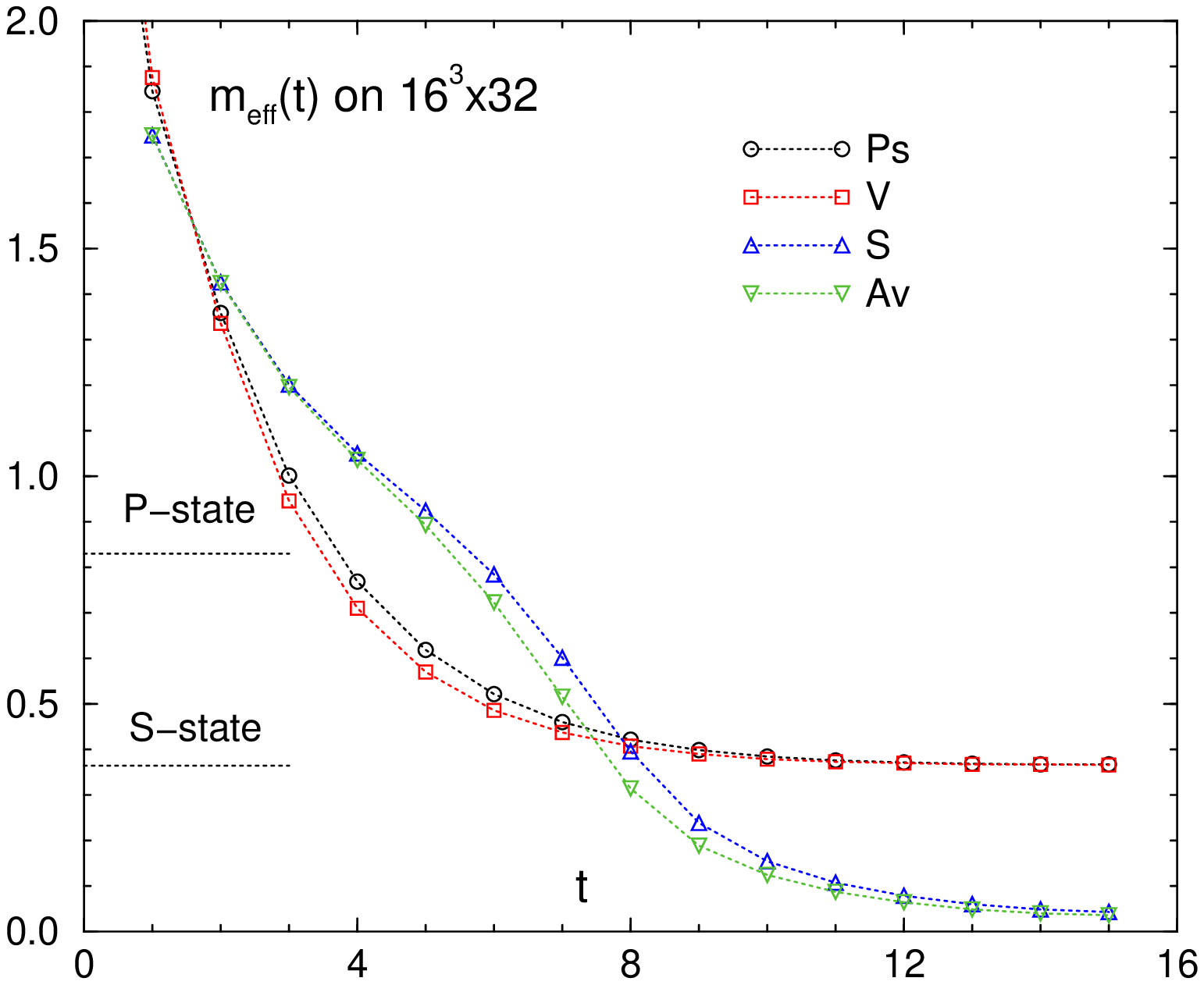}
 \includegraphics[width=.4\textwidth]{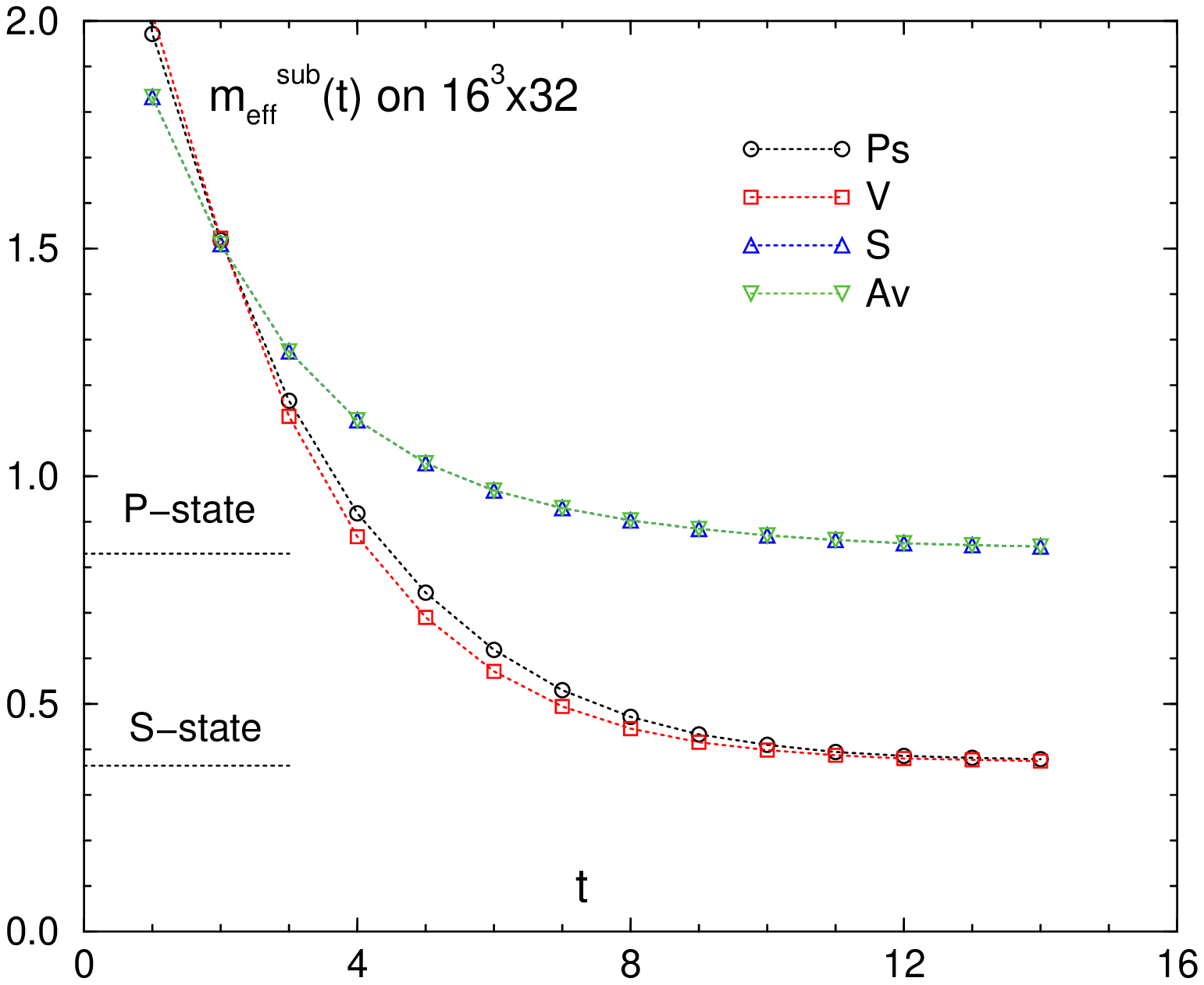}
\caption{(Left) Effective masses of meson-like correlators for each channel.
 These are free quark calculations on a $16^3\times 32$ lattice.}
\label{fig:free}
\end{center}
\end{figure}

Figure \ref{fig:free} (right panel) shows the effective mass 
$m_{\mbox{eff}}^{\mbox{sub}}(t)$, defined in Eq.~(\ref{eq:effmasssub}), 
from the free quark results for $N_s=16$.
The effective masses are equivalent to the usual effective mass shown 
in Fig.~\ref{fig:free} (left panel) except for the effects of the
constant mode.  
The expected energies of the lowest two quark states for the S-wave and
the P-wave states are 
shown in both figures of Fig.~\ref{fig:free}.
In contrast to the case of the usual effective masses, 
the subtracted effective masses
$m_{\mbox{eff}}^{\mbox{sub}}(t)$ approach the expected values even
in the P-wave states.  
The analysis to avoid the constant contribution works well
at least in the free quark case.

The method using the midpoint subtracted correlators can also be applied
to studies of spectral functions $\rho_\Gamma(\omega)$ by a modification
of the kernel,   
\begin{eqnarray}
\bar{C}(t)&=&\int_0^\infty d\omega \rho_\Gamma(\omega)K^{\mbox{sub}}
(\omega,t),\\
&&K^{\mbox{sub}}(\omega,t)=
\frac{2\sinh^2\left(\frac{\omega}{2}(\frac{N_t}{2}-t)\right)}
{\sinh{\left(\omega\frac{N_t}{2}\right)}}.
\label{eq:subkernel}
\end{eqnarray}
By using the alternative kernel $K^{\mbox{sub}}(\omega,t)$, it is
possible to extract the spectral function without the
contribution in $\omega\ll T$ from e.g. the Maximum Entropy Method. 

\section{Quenched QCD case at finite temperature}

\subsection{Lattice setup}
\label{sec:setup}

In this section we demonstrate effect of the constant mode and
the analysis to avoid it in quenched QCD at finite temperature.
Here we calculate the charmonium correlators, which are
defined like in Sect.~\ref{sec:freecorr} at the charm quark mass.

The gauge configurations have been generated by an standard plaquette
gauge action with a lattice gauge coupling constant, $\beta=6.10$ and a
bare anisotropy parameter $\gamma_G=3.2108$. 
The definition of the action and parameters
are the same as that adopted in Ref.~\cite{Matsufuru:2001cp}.
Lattice sizes are $20^3\times N_t$ where $N_t=160$ at $T=0$ and $N_t=32$,
26 and 20 at $T>0$. The lattice spacings are $1/a_s=2.030(13)$ GeV and
$1/a_t=8.12(5)$ GeV which are determined from the hadronic radius
$r_0=0.5$ fm.  
The physical volume size is about $(2\mbox{fm})^3$ and temperatures at
$N_t=32$, 26 and 20 are $T=0.88T_c$, $1.08T_c$ and $1.40T_c$
respectively. 

For the quark fields, we adopt an $O(a)$ improved Wilson quark action
with tadpole improved tree level clover coefficients.
Although the definition of the quark action is the
same as in Ref.~\cite{Matsufuru:2001cp}, 
we adopt a different choice of the Wilson
parameter $r=1$ to suppress effects of lattice artifacts
in higher states of charmonium \cite{Karsch:2003wy}.

\subsection{Finite temperature results}

\begin{figure}
\begin{center}
 \includegraphics[width=.48\textwidth]{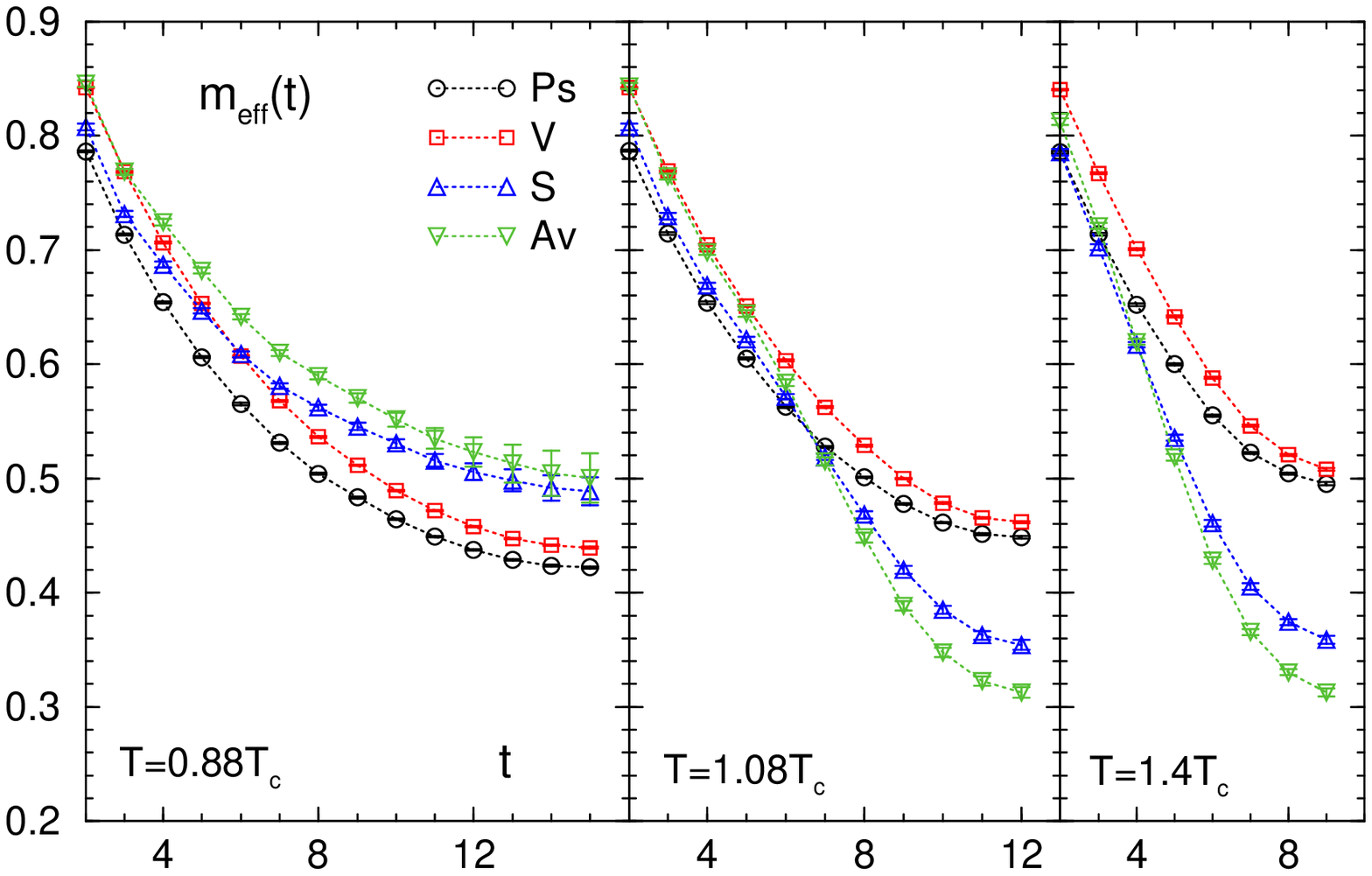}
 \includegraphics[width=.48\textwidth]{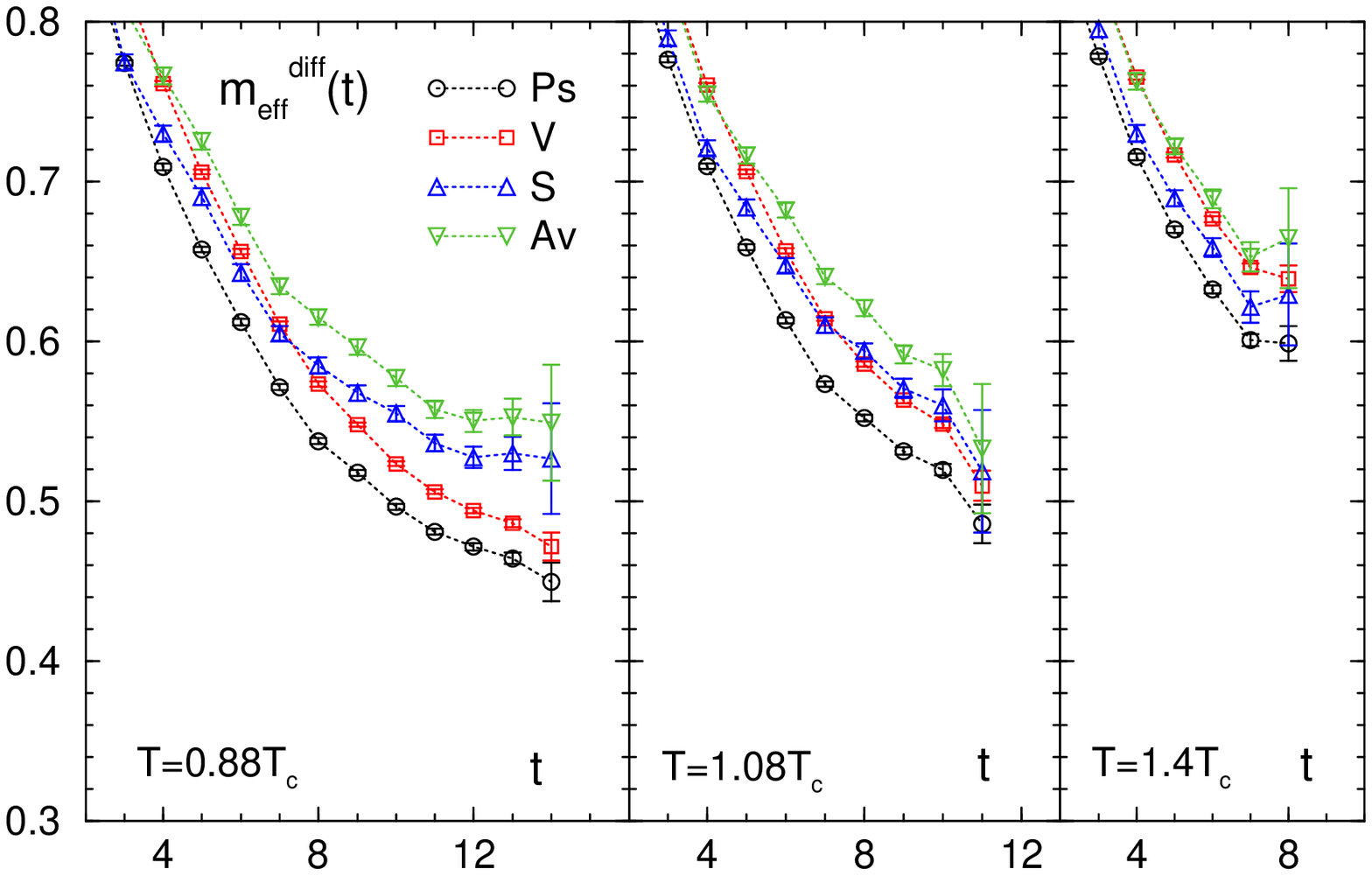}
\caption{(Left) Effective masses from usual charmonium correlators.
(Right) Effective masses from the midpoint subtracted correlators.
These are results with local operators for each channel 
in quenched QCD at finite temperature. }
\label{fig:quench1}
\end{center}
\end{figure}

First we show the usual effective masses, $m_{\mbox{eff}}(t)$ at finite
temperature in Fig.~\ref{fig:quench1} (Left panel). 
Since these are results obtained with local operators, there is no
plateau region at any temperature or channel we chose.
It can be expected from zero temperature calculation and the effective
masses show reasonable behavior below $T_c$, at which the charmonium
correlators does not show large difference from the zero temperature
results \cite{Umeda:2002vr,Datta:2003ww,Jakovac:2006sf}. 
Just above $T_c$, we see drastic changes in the P-wave state channels. 
The changes may cause the dissolution of $\chi_c$ states just above
$T_c$ as already reported in
Ref.~\cite{Datta:2003ww,Aarts:2006nr,Jakovac:2006sf}.
On the other hand, the S-wave states, i.e. the Ps and V channels, show
small changes up to  $1.4T_c$.  
This temperature dependences are shown in Fig.~\ref{fig:quench1} 
(left panel) for each channel. 
When we compare the effective masses at the same time
slices, we find no large change up to $1.4T_c$ in both channels as well. 
The results are consistent with previous lattice studies of charmonium
spectral functions 
\cite{Umeda:2002vr,Datta:2003ww,Asakawa:2003re,Aarts:2006nr,Jakovac:2006sf}, 
in which the spectral functions for the Ps and V channels in the 
deconfinement phase (but at not so high temperature) show a peak
structure around the $\eta_c$ and $J/\psi$ masses like the result at
zero temperature. 

\subsection{Midpoint subtracted correlator analysis}

Figure \ref{fig:quench1} (right panel) shows the results of
$m_{\mbox{eff}}^{\mbox{sub}}(t)$ in quenched QCD at finite temperature. 
In the P-wave states, the drastic changes of the usual
effective masses $m_{\mbox{eff}}(t)$ are absent in the effective masses
from the midpoint subtracted correlators $m_{\mbox{eff}}^{\mbox{sub}}(t)$.
Furthermore the similar behaviors of the effective masses hold till
$1.4T_c$ as in the case of the S-wave states.
The results indicate that the drastic changes of the usual correlators
(and effective masses) are caused only by the constant mode to
the correlators. 
The situation is very similar to the free quark case discussed in
Sect.~\ref{sec:free}.

\subsection{Results with spatially extended operators}

\begin{figure}
\begin{center}
 \includegraphics[width=.48\textwidth]{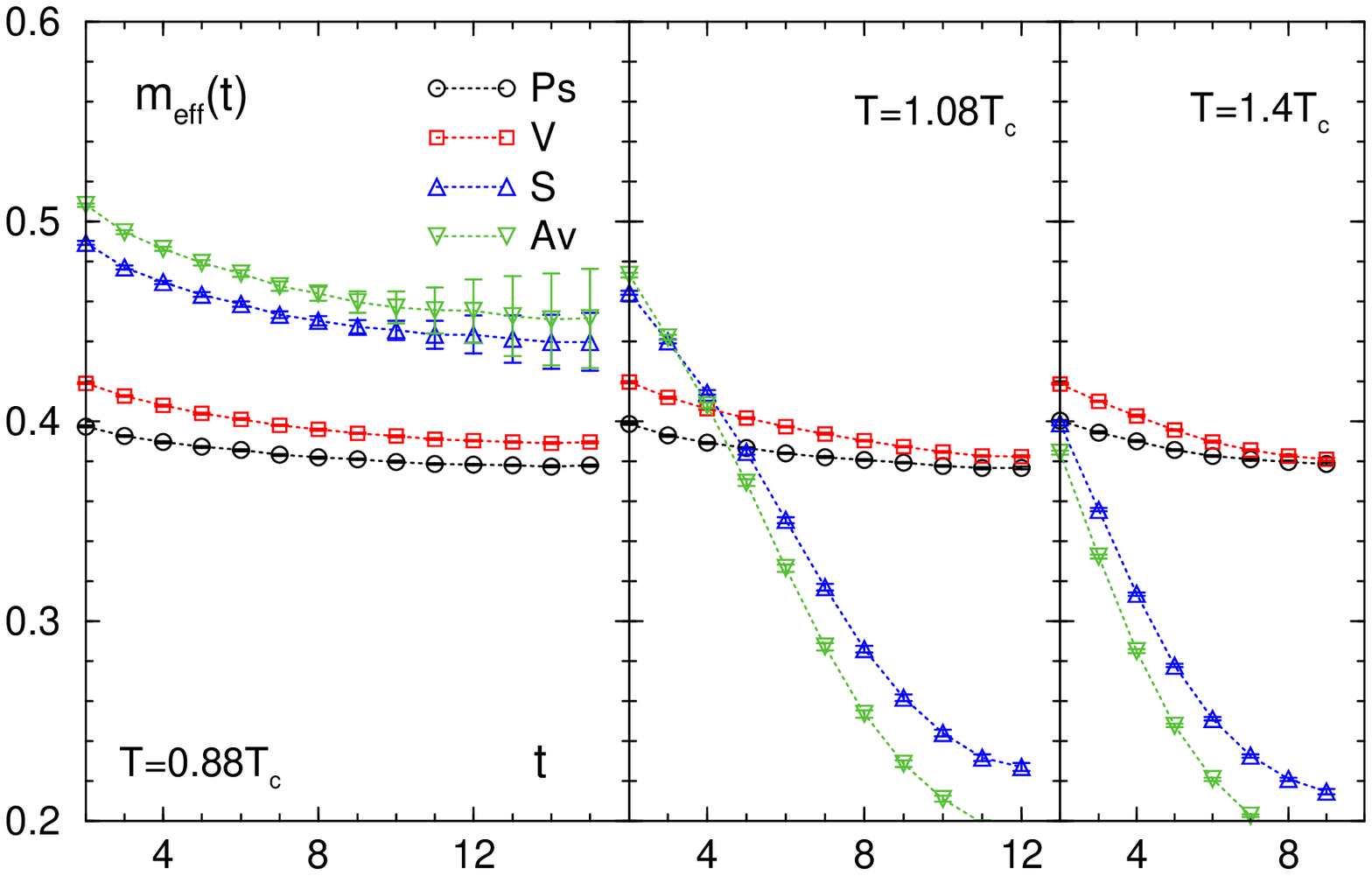}
 \includegraphics[width=.48\textwidth]{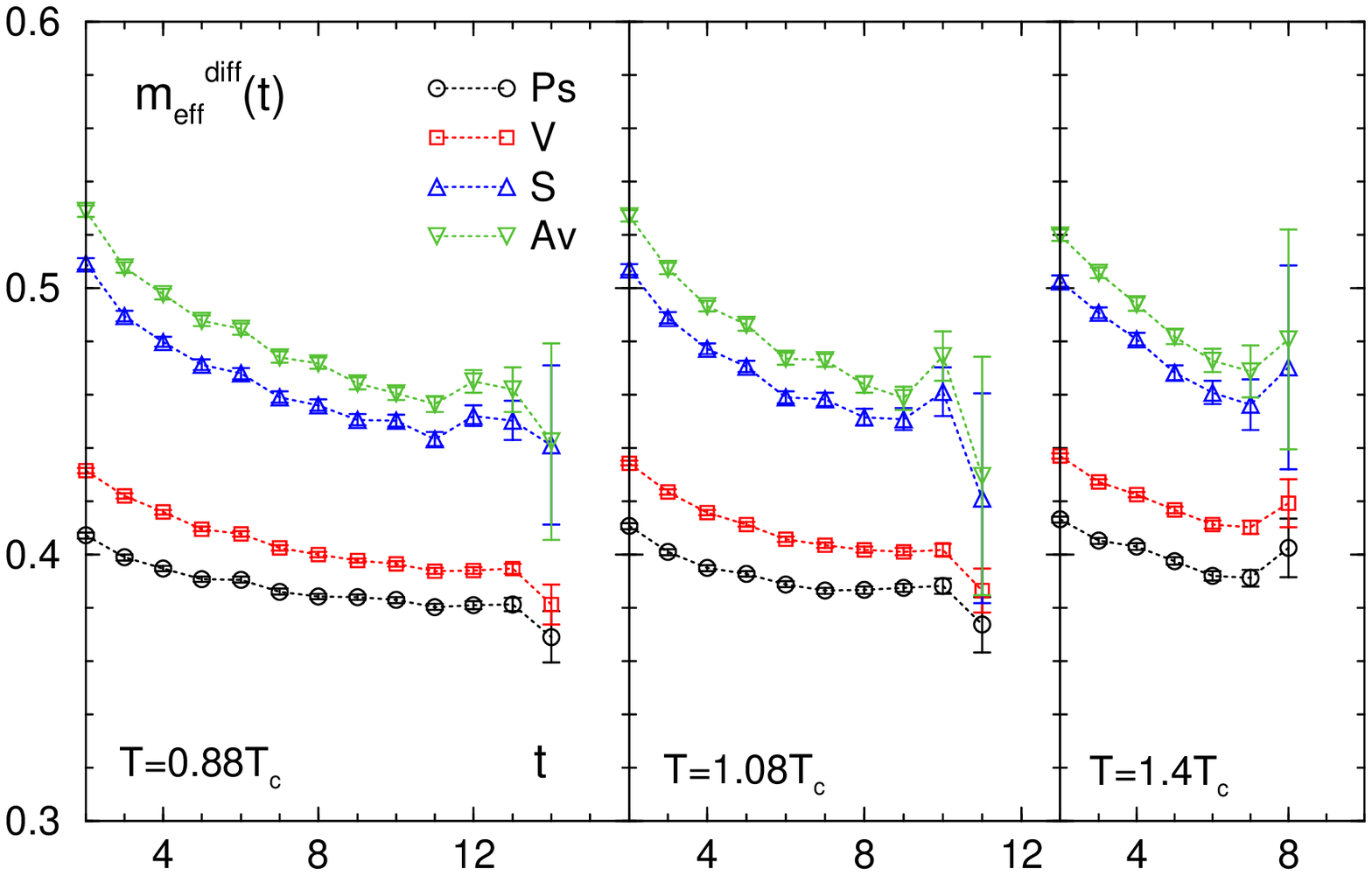}
\caption{(Left) Effective masses from usual charmonium correlators.
(Right) Effective masses from the midpoint subtracted correlators.
These are results with the spatially extended operators for each channel 
in quenched QCD at finite temperature. }
\label{fig:quench2}
\end{center}
\end{figure}

Finally we present results of meson correlators with spatially extended
operators.
The spatially extended operators are constructed with a smearing function
which yields a spatial distribution of quark (and anti-quark) source(s).
The spatially extended operators are defined by
$O_\Gamma(\vec{x},t)=\sum_{\vec{y}}\phi(\vec{y}) 
\bar{q}(\vec{x}-\vec{y},t)\Gamma q(\vec{x},t)$ 
with a smearing function $\phi(\vec{x})$ in Coulomb gauge.
In this calculation the spatially extended operators are adopted only in 
the source operators, the sink operators are local in any cases.
The smearing function is the same as that in Ref.~\cite{Umeda:2002vr},
i.e. $\phi(\vec{x})=\exp(-A|\vec{x}|^P)$ where $A$ and $P$ are parameters
determined by a matching with the charmonium wave function as
$A=0.2275$ and $P=1.258$.
Figure \ref{fig:quench2} (left panel) is the result for the usual
effective masses 
with the spatially extended operators. Other conditions are the same as 
in Fig.~\ref{fig:quench1}.
Below $T_c$, in contrast to the case of local operators the effective
masses reach a plateau due to the larger overlap with the lowest state.
Above $T_c$ the effective masses of the P-wave states
change more than in the case of local operators. 
Because the constant contribution is enhanced by the smearing function 
more than the other contributions in the case.
Figure \ref{fig:quench2} (right panel) is the same figure as
Fig.~\ref{fig:quench1},  
but with spatially extended operators.
We find similar behavior for both cases of operators.

\section{Conclusion}

Let us discuss details on $\chi_c$ states using the results of the
previous sections. In this study we find that the drastic changes of the
(usual) correlators in the P-wave states are caused by the constant
contribution,
while the correlators without the constant mode yield small change
till, at least, $1.4T_c$ even in the P-wave states. 
The changes are of similar size as that of the
S-wave states as one can see in Fig.~\ref{fig:quench1} and 
Fig.~\ref{fig:quench2}.

Although the constant mode coming from the scattering process is
important to investigate some transport phenomena, it does not affect 
the spectral function in the $\omega \gg T$ region for the charmonium 
correlators with zero spatial momentum.
Therefore the dissolution of $\chi_c$ states just above $T_c$ as
discussed in e.g. Ref.~\cite{Datta:2003ww,Aarts:2006nr,Jakovac:2006sf}
might be misleading. 
When there is the constant contribution, the hadronic state is no 
longer the lowest state. 
The situation provides some difficulties in the analysis
to reconstruct the spectral function in the higher energy part 
$\omega \gg T$ even if the MEM is applied \cite{Umeda:2005nw}. 

In principle the MEM analysis can extract the correct spectral function
even if the constant contribution exists, however it is difficult to
reproduce the non-lowest part of the spectral function correctly by the
MEM analysis at finite temperature with typical statistics.
In order to investigate the properties of $\chi_c$ states above $T_c$, a
careful analysis taking the constant contribution into account is
necessary. 
A correct MEM analysis of the correlator with and without the
constant contribution should give the same results, up to the 
low energy region $\omega\ll T$.
Such a comparison can be easily performed by the MEM analysis with
the alternative kernel of Eq.~(\ref{eq:subkernel}) for midpoint
subtracted correlators. 
Failure to do so signals unreliable MEM results.

\section*{Acknowledgments}

The simulations have been performed on supercomputers (NEC SX-5) at
the Research Center for Nuclear Physics (RCNP) at Osaka University and
(NEC SX-8) at the Yukawa Institute for Theoretical Physics (YITP) at
Kyoto University.  
This work has been authored under contract number
DE-AC02-98CH1-886 with the U.S. DOE
and Nos.19549001 with Grants-in-Aid of the Japanese MEXT.

\end{document}